\documentclass[conference]{IEEEtran}
\IEEEoverridecommandlockouts

\usepackage{cite}
\usepackage{amsmath,amssymb,amsfonts}
\usepackage{algorithmic}
\usepackage{graphicx}
\usepackage{textcomp}
\usepackage{xcolor}
\def\BibTeX{{\rm B\kern-.05em{\sc i\kern-.025em b}\kern-.08em
    T\kern-.1667em\lower.7ex\hbox{E}\kern-.125emX}}
\begin{document}

\title{Decentralized WebRCT P2P network using Kademlia\\}
% {\footnotesize \textsuperscript{*}Note: Sub-titles are not captured 

\author{\IEEEauthorblockN{Ryle Zhou}
% \IEEEauthorblockA{\textit{)} \\
\textit{Stanford Univerisity}\\
%  \\
rylezhou@stanford.edu}

\maketitle

\begin{abstract}
Web Real-Time Communication (WebRTC) is a new standard and industry effort that extends the web browsing model. For the first time, browsers are able to directly exchange real-time media with other browsers in a peer-to-peer fashion. Before WebRTC was introduced, it was cumbersome to build smooth chat and video applications, users often experience unstable connections, blurry videos, and unclear sounds. WebRTC's peer-to-peer communication paradigm establishes the real-time connection between browsers using the SIP(Session Initiation Protocol) Trapezoid. A wide set of protocols are bundled in WebRTC API, such as connection management, encoding/decoding negotiation, media control, selection and control, firewall and NAT element traversal, etc. However, almost all current WebRTC applications are using centralized signaling infrastructure which brings the problems of scalability, stability, and fault-tolerance. In this paper, I am presenting a decentralized architecture by introducing the Kademlia network into WebRTC to reduce the need for a centralized signaling service for WebRTC.
\end{abstract}

\begin{IEEEkeywords}
WebRTC, P2P, Kademlia Network, Distributed Systems
\end{IEEEkeywords}

\section{Introduction}
In 2020, COVID-19 has changed all of our lives. Video conferencing software are becoming essential to us. Because of the pandemic, cloud video conferencing services have never been more popular. With hundreds of millions of people working from home, the need of having smooth, high-quality voice and video chat systems that can support a large group of people staying online together at the same time has increased significantly. In this paper, I primarily explore the current progress with WebRTC and briefly talk about the combination of WebRTC and Kademlia Network with the purpose of having a decentralized way for WebRTC Peers to establish connections. 

\section{Fundamentals}

\subsection{WebRTC and its problem}
\subsubsection{WebRTC Infrastructure}
WebRTC \cite{b2} stands for Web Real-Time Communications.  For the first time, browsers are able to directly exchange real-time media with other browsers in a peer-to-peer fashion. WebRTC leverages a set of plugin-free APIs that can be used in both desktop and mobile browsers, and is progressively becoming supported by all major modern browser vendors. Previously, external plugins were required in order to achieve similar functionality as is offered by WebRTC. It includes a stack of protocols combined together(see Fig 1). In order to establish a peer-to-peer communication, web browsers must agree to begin communication, know how to locate one another, bypass security and firewall protections, and transmit all multimedia communications in real-time. In the web application scenario, the server can embed some JavaScript code in the HTML page it sends back to the client. Such code can interact with browsers through standard JavaScript APIs and with users through the user interface.
\begin{figure}[htbp]
\centerline{\includegraphics{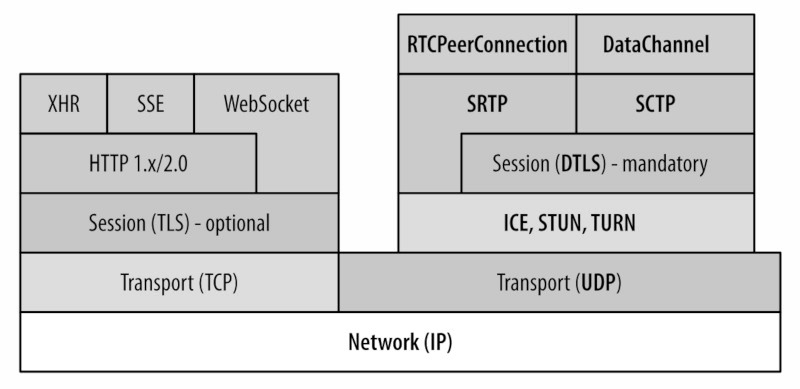}}
\caption{WebRTC protocol stack}
\label{fig}
\end{figure}

In WebRTC, both browsers are running a web application, which is downloaded from a different web server. Signaling messages are used to set up and terminate communications. They are transported by the HTTP or WebSocket protocol via web servers that can modify, translate, or manage them as needed. However, signalling is not implemented within WebRTC and it is a part that was left for developers to build based on different use cases(see Fig 2). 

\begin{figure}[htbp]
\centerline{\includegraphics[scale=.2]{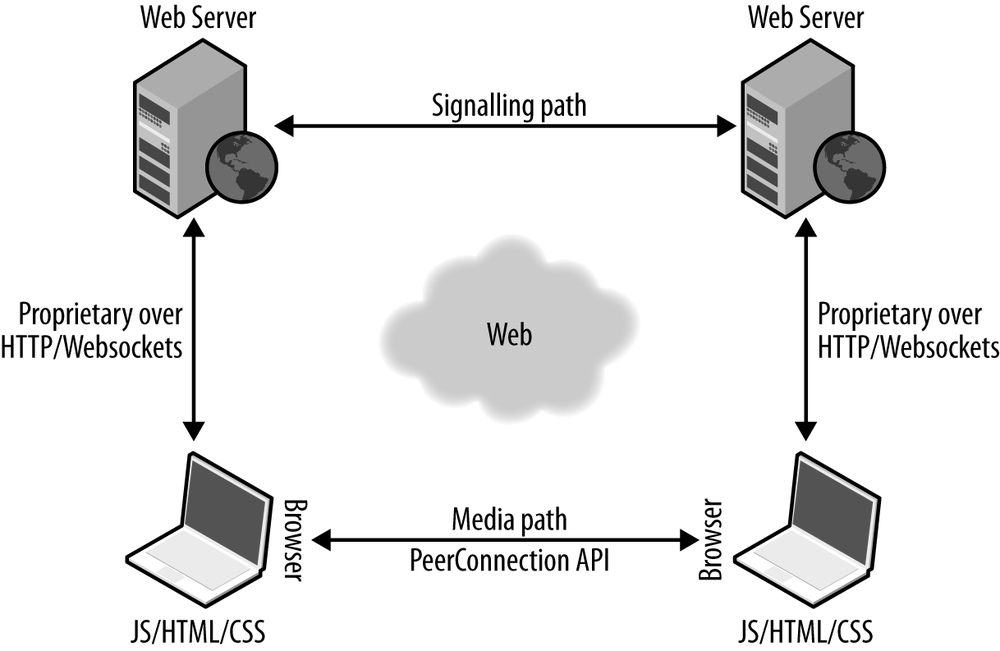}}
\caption{WebRTC Infrastructure}
\label{fig}
\end{figure}

\subsubsection{Data Channel}
The \emph{DataChannel} API is designed to provide a generic transport service allowing web browsers to exchange generic data in a bidirectional peer-to-peer fashion.
The encapsulation of SCTP over DTLS over UDP together with ICE provides a NAT traversal solution, as well as confidentiality, source authentication, and integrity protected transfers. Moreover, this solution allows the data transport to interwork smoothly with the parallel media transports, and both can potentially also share a single transport-layer port number. SCTP has been chosen since it natively supports multiple streams with either reliable or partially reliable delivery modes. It provides the possibility of opening several independent streams within an SCTP association towards a peering SCTP endpoint. Each stream actually represents a unidirectional logical channel providing the notion of in-sequence delivery. A message sequence can be sent either ordered or unordered. The message delivery order is preserved only for all ordered messages sent on the same stream. However, the \emph{DataChannel} API has been designed to be bidirectional, which means that each \emph{DataChannel} is composed as a bundle of an incoming and an outgoing SCTP stream.

The \emph{DataChannel} setup is carried out (i.e., the SCTP association is created) when the CreateDataChannel() function is called for the first time on an instantiated PeerConnection object. Each subsequent call to the \emph{CreateDataChannel()} function just creates a new \emph{DataChannel} within the existing SCTP association.

\subsubsection{Peer Connection}
The peer-to-peer connectivity is handled by the \emph{RTCPeerConnection} interface. A \emph{RTCPeerConnection} allows two users to communicate directly, browser to browser. It then represents an association with a remote peer, which is usually another instance of the same JavaScript application running at the remote end. Communications are coordinated via a signaling channel provided by scripting code in the page via the web server, e.g., using XMLHttpRequest or WebSocket. Once a peer connection is established, media streams can be sent directly to the remote browser. Media capture devices includes video cameras and microphones, but also screen capturing "devices". For cameras and microphones, the  \emph{navigator.mediaDevices.getUserMedia()} handles the \emph{MediaStreams}.
For example, Peer 1 creates an offer and initiates signalling with STUN/TURN server, and as a result, Peer 1 receives the ICE candidates. Both the offer and the ICE candidates are sent to peer 2 through the signaling channel. As soon as Peer 2 receives the offer, it creates an answer and performs the same process and sends its connectivity information through the same signaling channel back to Peer 1. After the signaling is complete, both peers have all connectivity information: answer, offer and both ICE candidates.

\subsubsection{STUN AND TURN}
The Session Traversal Utilities for NAT (STUN) protocol (RFC5389) allows a host application to discover the presence of a network address translator on the network, and in such a case to obtain the allocated public IP and port tuple for the current connection. To do so, the protocol requires assistance from a configured, third-party STUN server that must reside on the public network.

The Traversal Using Relays around NAT (TURN) protocol (RFC5766) allows a host behind a NAT to obtain a public IP address and port from a relay server residing on the public Internet. Thanks to the relayed transport address, the host can then receive media from any peer that can send packets to the public Internet.

The \emph{RTCPeerConnection} mechanism uses the ICE protocol together with the STUN and TURN servers to let UDP-based media streams traverse NAT boxes and firewalls. ICE allows the browsers to discover enough information about the topology of the network where they are deployed to find the best exploitable communication path. Using ICE also provides a security measure, as it prevents untrusted web pages and applications from sending data to hosts that are not expecting to receive them.

Each signaling message is fed into the receiving \emph{RTCPeerConnection} upon arrival. The APIs send signaling messages that most applications will treat as opaque blobs, but which must be transferred securely and efficiently to the other peer by the web application via the web server.

\subsubsection{The problem with ICE, STUN AND TURN}
STUN servers live on the public internet and have one simple task: check the IP:port address of an incoming request (from an application running behind a NAT) and send that address back as a response. In other words, the application uses a STUN server to discover its IP + port from a public perspective. This process enables a WebRTC peer to get a publicly accessible address for itself, and then pass that on to another peer via a signaling mechanism, in order to set up a direct link(
see Fig 3).

\begin{figure}[htbp]
\centerline{\includegraphics[scale=.2]{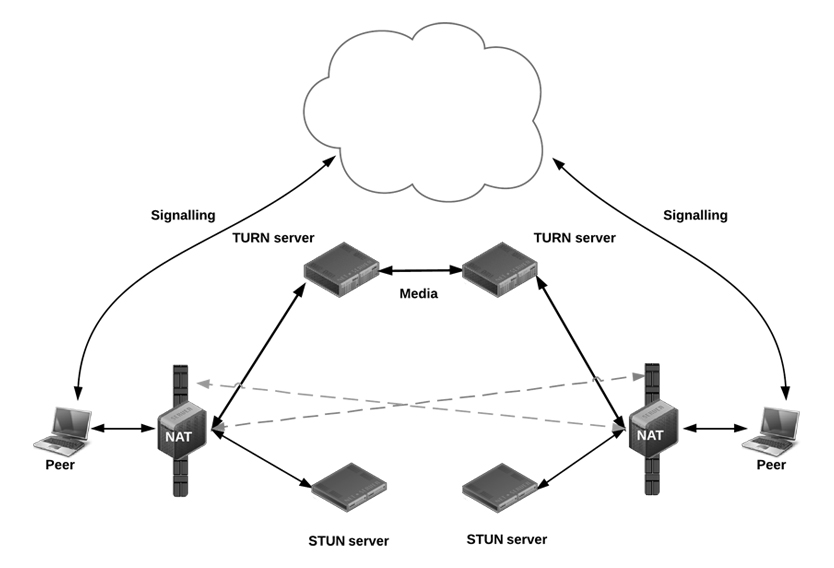}}
\caption{WebRTC Peer Connection}
\label{fig}
\end{figure}
RTCPeerConnection tries to set up direct communication between peers over UDP. If that fails, RTCPeerConnection resorts to TCP. If that fails, TURN servers can be used as a fallback, relaying data between endpoints. TURN servers have public addresses, so they can be contacted by peers even if the peers are behind firewalls or proxies. 
WebRTC-based P2P networks always require two components PeerConnction: A signaling mechanism to exchange the offer/answer messages and the ICE candidates, and a STUN/TURN server to establish connections from/to peers behind NATs. As a result, WebRTC is behind centralized distributed systems that might be signal point of failures, downtime, instability and other unreliable problems. 

\subsection{Kademlia Network}
Kademlia\cite{b1} is a distributed hash table for decentralized P2P networks. It specifies the structure of the network and the exchange of information through node lookups. Kademlia nodes communicate among themselves using UDP. A virtual or overlay network is formed by the participant nodes. Each node is identified by a number or node ID. The node ID serves not only as identification, but the Kademlia algorithm uses the node ID to locate values (usually file hashes or keywords). In fact, the node ID provides a direct map to file hashes and that node stores information on where to obtain the file or resource. In Kademlia, XOR metric is introduced to define a distance between two nodes. The XOR distance is the bitwise exclusive OR on the peers’ identifiers interpreted as an integer(see Fig 4). 
\begin{figure}[htbp]
\centerline{\includegraphics[scale=.6]{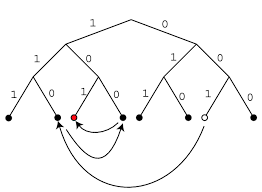}}
\caption{Kademlia XOR Metric}
\label{fig}
\end{figure}

When searching for some value, the algorithm needs to know the associated key and explores the network in several steps. Each step will find nodes that are closer to the key until the contacted node returns the value or no more closer nodes are found. This is very efficient: like many other DHTs, Kademlia contacts only $O(\log(n))$ nodes during the search out of a total of n nodes in the system. For example, if there are 10,000,000 Kademlia nodes, only about 20 hops would be necessary at most for communication with any subset of nodes. Kademlia based networks are highly resistant to denial of service attacks and the loss of a group of nodes as the protocol simply routes around the unavailable nodes.

Further advantages are found particularly in the decentralized structure, which increases the resistance against a denial-of-service attack. Even if a whole set of nodes is flooded, this will have limited effect on network availability, since the network will recover itself by knitting the network around these "holes".

Another advantage of Kademlia is that the protocol naturally prefers long-lived nodes over newer entrants. The process of joining a Kademlia network requires discovery of only one peer, whereby the node then broadcasts its appearance. The initiator then collects the NodeID from each response and adds it to its own peer table.

\subsubsection{Kademlia NodeID}
Kademlia treats each node on a network as a leaf on a binary tree. Generally, each Kademlia node has a 160-bit NodeID (SHA-1), and its position is determined by the shortest unique prefix of its ID.
To assign key-value pairs to particular nodes, Kademlia relies on a notion of distance between two identifiers. Given two 160-bit identifiers, x and y, Kademlia defines the distance between them as the XOR.
From a node point of view, the tree is divided into series of successive sub-trees where the 160th sub-tree contains the individual node. The Kademlia protocol ensures that each node knows of at least one node on each of its sub-trees. With this guarantee, a node can locate any other node by its ID.

\subsubsection{Routing Table and K-buckets}
The routing table is a binary tree whose leaves are k buckets. The structure of the Kademlia routing table is such that nodes maintain detailed knowledge of the address space closest to them, and exponentially decreasing knowledge of more distant address space.
The symmetry is useful since it means that each of these closest contacts will be maintaining detailed knowledge of a similar part of the address space, rather than a remote part.
K-buckets are a list of routing addresses of other nodes in the network, which are maintained by each node and contain the IP address, port, and NodeID for peer participants in the system. They prefer the longest-lived nodes, which means that one cannot overtake a node’s routing state by flooding the system with new nodes.
The routing table size is asymptotically bounded by O(log(n/k)) where n is the actual number of nodes in the network and k is the bucket size, so larger bucket implementations slightly reduce the total number of buckets in the routing table.

The Kademlia protocol consists of four Remote Procedure Calls :\newline
PING: probes a node to see if it’s online. \newline
STORE: instructs a node to store a key-value pair.\newline
FIND\_NODE: returns information about the k nodes closest to the target ID. \newline
FIND\_VALUE: similar to the FIND\_NODE , but if the recipient has received a STORE for the given key, it just returns the stored value.

\section{Decentralized WebRTC with Kademlia}
From the above section, I found that the ICE, STUN/TURN part of WebRTC is used to deal with NAT traversal in a centralized way. In this section, I bring up the notion that connecting peers in WebRTC through a Kademlia network. Although, the actual network might still be in a server network owned by some companies. However, using Kademila can create this WebRTC service in a decentralized way. 
To join the network, local Peer A make a connection through WebSockets to a node(system configuration, this node is defined as a long-lived node) in Kademlia network and sends a FIND\_NODE request that returns the k closest node to the remote Peer B. Here, the node nominated in Kademlia network acts as the signalling channel. 
The implementation of Kademlia in AWS was actually challenging to me. I encountered many bugs and a lot of details could go wrong. If implemented correctly, STUN/TURN server can be replaced by the Kademlia network. And theoretically, through iterative routing, the time to find the target node will be fast. 
To evaluate this model, I thought about a few metrics to measure: \newline
\begin{itemize}
\item RTCConnection time for each call session to establish measured for 100 times
\item Failure Rate (here failure means no connection was established)
\item Call Session time that had no dropped connection
\end{itemize}

\section{Conclusion and Future work}
In this paper, I explored WebRTC work process, Kademlia Network and the combination of the two technology in order to create a decentralized chat service in a distributed system. Because of the centralization nature of the  of ICE, STUN/TURN server and signaling channel, I thought it would be a problem for to have a reliable tool for people. The purpose is to reduce the risk of single point of failure and establish a more reliable and available system. There are a lot of technical challenges involved such as, deploying Kademlia network on cloud, signalling configuration and the front end implementation etc. Here, I did not explore how the security of the network, which is a big concern among the public right now. \newline
My future work includes continue figuring out how well Kademlia can work with WebRTC, possibly using a load balancer to further test scalability. The possibility of extending WebRTC to mobile devices is also worth taking into consideration.

\vspace{12pt}
\color{red}
% IEEE conference templates contain guidance text for composing and formatting conference papers. Please ensure that all template text is removed from your conference paper prior to submission to the conference. Failure to remove the template text from your paper may result in your paper not being published.

\end{document}